\documentclass[sigplan,10pt]{acmart}
\renewcommand\footnotetextcopyrightpermission[1]{}
\AtBeginDocument{%
  
}

\usepackage{booktabs}
\usepackage{fontawesome}
\usepackage{xcolor}
\usepackage{colortbl}
\usepackage{pdflscape}  
\usepackage{adjustbox}  
\usepackage{multirow}
\usepackage{array}
\usepackage{makecell}
\usepackage{graphicx}
\usepackage{xspace}
\usepackage{xurl}
\usepackage{bbding}
\usepackage{caption}
\usepackage{subcaption}

\usepackage[newfloat,frozencache,cachedir=.]{minted}

\usepackage{caption}
\newenvironment{code}{\captionsetup{type=listing}}{}
\SetupFloatingEnvironment{listing}{name=Code}

\definecolor{yescolor} {HTML}{2e7d32}   
\definecolor{partcolor}{HTML}{e65100}   
\definecolor{nocolor}  {HTML}{c62828}   
\definecolor{scenicbg} {HTML}{e8f0fe}   

\newcommand{\Yes} {\textcolor{yescolor} {\faCheckSquare}}
\newcommand{\Part}{\textcolor{partcolor}{\faPlusSquareO}}
\newcommand{\No}  {\textcolor{nocolor}  {\faTimesCircle}}

\usepackage{tikz}
\usepackage{circledtext}

\newcommand*{\RobustCircled}[2][1pt]{%
    \tikz[baseline=(char.base), scale=0.75]{%
        \node[shape=circle, 
              draw=black,     
              line width=0.5pt, 
              inner sep=#1] (char) {#2};%
    }%
}

\begin{document}
\pagestyle{plain}
\title[]{SCENIC: Stream Computation-Enhanced SmartNIC}

\author{Benjamin Ramhorst}
\authornote{Equal contribution.}
\email{bramhorst@ethz.ch}
\affiliation{
\institution{ETH Zurich}
\city{Zurich}
\country{Switzerland}
}

\author{Maximilian J. Heer}
\authornotemark[1]
\email{hmaximili@ethz.ch}
\affiliation{
\institution{ETH Zurich}
\city{Zurich}
\country{Switzerland}
}

\author{Luhao Liu}
\email{luhliu@ethz.ch}
\affiliation{
\institution{ETH Zurich}
\city{Zurich}
\country{Switzerland}
}

\author{Heejae Kim}
\email{adpp00@snu.ac.kr}
\affiliation{
\institution{Seoul National University}
\city{Seoul}
\country{Korea}
}

\author{Jonas Dann}
\email{jodann@ethz.ch}
\affiliation{
\institution{ETH Zurich}
\city{Zurich}
\country{Switzerland}
}

\author{Jin-Soo Kim}
\email{jinsoo.kim@snu.ac.kr}
\affiliation{
\institution{Seoul National University}
\city{Seoul}
\country{Korea}
}

\author{Gustavo Alonso}
\email{alonso@inf.ethz.ch}
\affiliation{
\institution{ETH Zurich}
\city{Zurich}
\country{Switzerland}
}

\renewcommand{\shortauthors}{}

\begin{abstract}
Although modern, AI-centric datacenters heavily rely on SmartNICs, existing devices impose a hard trade-off. Commercial SmartNICs provide high bandwidth and easy software integration, but offer limited support for customization and data processing offload. In contrast, research SmartNICs often suffer from low bandwidth, limited functionality, and poor software compatibility - to the point that many are not actual NICs in a technical sense. This gap can be closed by treating the NIC datapath as a first-class stream computation substrate with shared hardware/software abstractions for a tight co-design of infrastructure and applications. To demonstrate this, we introduce SCENIC, an open-source datacenter SmartNIC. SCENIC implements a 200G network datapath over offloaded TCP/IP and RDMA stacks, as well as a fallback path for processing arbitrary network traffic. On top of the network logic, SCENIC combines on-datapath Stream Compute Units (SCUs) for data processing and embedded ARM cores for flexible control path manipulation with direct access to GPUs and SSDs. SCENIC is fully integrated with the OS, exposing native Linux network and RDMA verb interfaces, making the programmable datapath transparent to existing applications while enabling control of, e.g., user-defined offloads and programmable congestion control. SCENIC’s performance matches commercial platforms, and we show its versatility through several use cases such as offloaded collective communication and network-to-GPU hash-based data partitioning.

\end{abstract}

\maketitle

\section{Introduction}
With the ever-increasing scale of modern applications and improvements in compute efficiency through hardware specialization~\cite{cpu-decline-cacm, seattle-db-report-cacm}, computer systems are becoming increasingly constrained by network performance~\cite{network-bottleneck}. Following current trends in compute and network bandwidth scaling, estimates indicate that distributed communication will make up between 50\% and 75\% of future training run-time~\cite{network-bottleneck-2}. At the same time, around 25-30\% of CPU cycles in datacenters are spent on infrastructure tasks, often referred to as the "datacenter tax"~\cite{datacenter-tax}. To tackle these problems, recent focus has shifted to augment and optimize networking with compute through Smart Network Interface Cards (SmartNICs). SmartNICs, in addition to implementing the packet processing pipeline, also offload various steps of the data processing and compute pipeline. This includes, for example, network virtualization functions, storage access, network security and transport protocols, as shown by large-scale deployments of AWS Nitro~\cite{aws-nitro}, Microsoft AccelNet~\cite{accelNet} and Alibaba CIPU~\cite{alibab_cipu}. Furthermore, off-the-shelf SmartNICs, such as NVIDIA Bluefield~\cite{bluefield3}, AMD Pensando~\cite{pensandoElba} and Broadcom Stingray~\cite{stingray}, have become the backbone of modern ML systems, forming a scale-out network for thousands of GPUs~\cite{oracle2025zettascale10}. In research, SmartNICs have been used to explore in-network compute for ML systems~\cite{guo_ml_smartnic, conspirator}, storage offloads~\cite{smartnic_storage_zeke, stream_dedup}, databases~\cite{honeycomb, alnico, skv}, and security~\cite{zhao_smartnic_security, panda_smartnic_security}.

Despite their high network bandwidth and ease-of-use, the very nature of closed-source and hardened commercial SmartNICs hinders novel research and adaptation to modern workloads. For example, next-generation protocols (e.g. UltraEthernet)~\cite{ultraethernet_principles, google_falcon, amazon_srd} and congestion control algorithms~\cite{hpcc, revisting_cc_ethernet} are explored primarily through simulation due to the limited customizability of commercial NICs. Additionally, these SmartNICs often suffer from poor offload performance, as they are typically implemented using off-path Arm cores with high memory access latency~\cite{off_path_compute_2} or on-path RISC-V cores with limited single-thread performance~\cite{off_path_compute_1}. 

There has also been an increasing amount of interest in SmartNICs from the research community, with projects exploring many variations of the idea~\cite{corundum, opennic, recoNIC, hXDP}. However, most of these systems are limited in bandwidth (100G and often less), lack support for transport protocol offloading (e.g., RDMA, TCP/IP), have no native OS integration (e.g., through Linux \texttt{netdev} or \texttt{ibv\_device}), and have limited or no integration with GPUs or SSDs like commercial SmartNICs do today. With some exceptions, they are not maintained, being just prototypes to demonstrate an idea or the potential to offload some functionality to NICs. 

In this paper, we present SCENIC, an FPGA-based SmartNIC with end-to-end system integration, designed to support in-network data processing and enabling full customization. SCENIC exploits the streaming nature of network traffic by introducing the notion of reprogrammable Stream Computation Units (SCU) that can be assigned to process network flows in arbitrary ways. The SCUs can be utilized with any of the  offloaded network stacks (RDMA, TCP/IP) as well as in combination with collective communication primitives. SCENIC supports up to 200G bandwidth with native OS integration through Linux \texttt{netdev} and \texttt{ibv\_device}. The resulting system is similar to those used at scale by Microsoft in Azure cloud~\cite{microsoft_fpga_at_scale} or NVIDIA to manage communication for AI accelerators~\cite{nvidia_lpx}, but with SCENIC being an open-source project providing higher customization possibilities across the entire stack\footnote{GitHub repository: https://github.com/fpgasystems/SCENIC}. SCENIC builds on well-established, open-source projects (shell~\cite{coyote, coyotev2}, network stacks~\cite{roce_balboa, fpga-network-stack-github}, communication libraries~\cite{accl+}, and applications) that had to be redesigned for higher bandwidth, new FPGA architectures and native datacenter compatibility. As such, it offers the possibility of modifying or replacing all of its components, tailoring the system for individual applications or research use cases. SCENIC's contributions include:
\begin{itemize}
    \item A high-performance network datapath with offloads for common protocols (RDMA, TCP/IP), as well as a fallback path for processing arbitrary traffic. 
    \item Support for multiple parallel, isolated applications on the network datapath. Uniquely, SCENIC enables the deployment of offloaded applications both in programmable logic, similar to FPGA-based SmartNICs, and in on-chip Arm cores, similar to commercial DPUs.
    \item A virtual memory model which allows offloaded applications to access both NVIDIA and AMD GPUs, as well as conventional SSDs.
    \item Native integration with Linux \texttt{netdev} and \texttt{ibv\_device}, exposing SCENIC as a standard NIC and ensuring compatibility with existing networked applications.
    \item A thorough evaluation covering throughput, latency, GPU/SSD integration, and fairness, demonstrating performance comparable to commercial 200G devices.
    \item A demonstration of SCENIC's capabilities through two use cases: SmartNIC-offloaded collective communication with performance comparable to OpenMPI, and hash-based data partitioning for multi-GPU execution of database operators, with a 6.7x improvement over the CPU baseline.
\end{itemize}

\section{Related Work}
Throughout this paper, we adopt the definition of a \textit{NIC} from~\cite{ocp_nic_definition}: a PCIe device with (i) a physical network interface (e.g., Ethernet or InfiniBand), (ii) a DMA engine for host-memory packet transfers, and (iii) an MSI-X interrupt interface for completions. To preserve compatibility with decades of existing work and avoid software rewrites, we further require a NIC to provide a host driver compatible with standard networking frameworks (e.g., Linux \texttt{netdev}, \texttt{ibv\_device}, or DPDK). \textit{SmartNICs} extend conventional NICs by integrating programmable compute on the card~\cite{smartnic-survey-2, ajayi2024chronological, smartnic-survey-1, doring2021smartnics, elizalde2025security}, while also ensuring compatibility with existing networking frameworks. In the following, we discuss related platforms and compare them to SCENIC. We explicitly distinguish between feature-complete SmartNICs and application-specific networking platforms and components.

\begin{table*}[t]
\footnotesize
\centering
\caption{%
  Comparison of SmartNICs and related projects.
  \Yes~supported;\;
  \Part~partially supported;\;
  \No~not supported.\
}
\label{tab:nic-comparison}
\setlength{\tabcolsep}{2pt}
\renewcommand{\arraystretch}{1.2}
\resizebox{\textwidth}{!}{%
%
%
\newcolumntype{R}{>{\raggedright\bfseries\arraybackslash}p{3.4cm}}
\newcolumntype{C}{>{\centering\arraybackslash}p{1.05cm}}
%
%
\begin{tabular}{R C C C C C C C C C C C}
\toprule
%
%
\textbf{System}
  & \rotatebox{90}{Network Speed\hspace{6pt}}
  & \rotatebox{90}{TCP Offload\hspace{6pt}}
  & \rotatebox{90}{RDMA Offload\hspace{6pt}}
  & \rotatebox{90}{\texttt{netdev} driver\hspace{6pt}}
  & \rotatebox{90}{\texttt{ibv\_device} impl.\hspace{6pt}}
  & \rotatebox{90}{ARM Cores\hspace{6pt}}
  & \rotatebox{90}{Stream Compute\hspace{6pt}}
  & \rotatebox{90}{PCC\hspace{6pt}}
  & \rotatebox{90}{Open-Source\hspace{6pt}}
  & \rotatebox{90}{Direct-to-GPU\hspace{6pt}}
  & \rotatebox{90}{Direct-to-Storage\hspace{6pt}}
\\
\midrule
%
%
\rowcolor{scenicbg}
SCENIC (ours)
  & 200\,G  & \Yes  & \Yes  & \Yes  & \Yes
  & \Yes    & \Yes  & \Yes  & \Yes
  & \Yes    & \Yes  
\\
%
%
\midrule
\multicolumn{12}{l}{\small\textit{Group 1: Academic FPGA-based NICs}}
\\[1pt]
Corundum~\cite{corundum}
  & 100\,G & \No   & \No & \Yes  & \No
  & \No        & \Yes   & \No   & \Yes
  & \No        & \No  
\\
OpenNIC~\cite{opennic}
  & 100\,G  & \No   & \No   & \Yes  & \No
  & \No     & \Yes   & \No   & \Yes
  & \No     & \No   
\\
RecoNIC~\cite{recoNIC}
  & 100\,G  & \No   & \Yes  & \Yes & \No
  & \No     & \Yes   & \No   & \Part
  & \No     & \No   
\\
hXDP~\cite{hXDP}
  & 40\,G & \No  & \No   & \Yes  & \No
  & \No       & \No  & \No   & \Yes
  & \No       & \No  
\\
ZeroNIC~\cite{zero_nic}
  & 100\,G & \Part  & \No   & \Yes  & \Yes
  & \No       & \No  & \Part   & \No
  & \Yes       & \No  
\\

%
%
\midrule
\multicolumn{12}{l}{\small\textit{Group 2: Commercial NICs / DPUs}}
\\[1pt]
Broadcom Stingray~\cite{stingray}
  & 25/100\,G & \Part & \Yes   & \Yes  & \Yes
  & \Yes      & \No   & \No   & \No
  & \Yes       & \Yes  
\\
Intel IPU E2000~\cite{intelIPU}
  & 200\,G  & \Part & \Yes   & \Yes  & \Yes
  & \Yes    & \Part   & \Part & \No
  & \Yes   & \Yes  
\\
NVIDIA BlueField-3 B3220~\cite{bluefield3}
  & 2x200\,G & \No  & \Yes  & \Yes  & \Yes
  & \Yes         & \Part   & \Yes  & \No
  & \Yes         & \Yes  
\\
AMD Pensando Elba~\cite{pensandoElba}
  & 2x200\,G & \Part  & \Yes  & \Yes  & \Yes
  & \Yes         & \Part    & \Yes  & \No
  & \Yes         & \Yes 
\\
Mango BoostX~\cite{mangoboost}
  & $\leq$400\,G & \Yes & \Yes  & \Yes & \Yes
  & \Yes          & \Yes  & \Yes & \No
  & \Yes         & \Yes   
\\

\bottomrule
\end{tabular}%
}
\end{table*}

\subsection{Feature-complete SmartNICs}

The central role of networking in cloud and datacenter computing 
has led to a number of academic SmartNICs based on FPGAs~(\autoref{tab:nic-comparison}).
Corundum~\cite{corundum} and OpenNIC~\cite{opennic} represent relatively early designs 
that focus on pure network connectivity with added compute for control plane operations. Consequently, neither provides offloaded network stacks (RDMA or TCP/IP) nor integration with GPUs or SSDs.
RecoNIC~\cite{recoNIC} extends OpenNIC with a hardware-offloaded RDMA stack enabling direct,
zero-copy data transfer over RoCEv2.
hXDP~\cite{hXDP}  focuses on offloading Linux XDP/eBPF
programs on FPGAs hardware, representing a programming model orthogonal to SCENIC. It builds on top of NetFPGA~\cite{netfpga}, which provides a reference design for the network driver and the hardware implementation. Similar to OpenNIC and Corundum, it does not include a complete network stack nor integration with other devices (GPUs, SSDs).
Additionally, all of the aforementioned platforms are bandwidth-bound to 100G or less. 
The same is true for ZeroNIC~\cite{zero_nic}, which, similar to SCENIC, aims at datacenter compatibility through support for GPUs and exposure as \texttt{netdev} and \texttt{ibv\_device}. ZeroNIC's programmability lies in its flexible, software-defined control plane, rather than in on-NIC data processing capabilities. 

Turning to commercial platforms, a similar analysis can be made. 
The Broadcom Stingray~\cite{stingray} targets general-purpose infrastructure offload with ARM cores and hardware accelerators for crypto, RAID, and storage, but predates the current generation of DPU platforms in both bandwidth and programmability. Intel IPU E2000~\cite{intelIPU}, NVIDIA BlueField-3~\cite{bluefield3}, and AMD Pensando Elba~\cite{pensandoElba} combine hardware RoCE engines, ARM SoC cores, IPSec and virtualization offloads, storage acceleration and GPU-centric networking at up to 2×200G. All three provide programmable datapath capabilities through P4-programmable ARM cores (E2000), multi-threaded RISC-V cores (BF3), and P4-programmable match-action pipelines (Pensando), which are generally more constrained than a fully customizable datapath on an FPGA. The FPGA-based MangoBoost BoostX~\cite{mangoboost} explores a more programmable variant at up to 400G with user-specified on-FPGA logic. However, MangoBoost's own documentation states that it may contain forward-looking statements which are subject to change, so some features reported in \autoref{tab:nic-comparison} may not actually reflect the shipped product.
All of the commercial platforms are closed-source, proprietary products that limit customization and research use~\cite{smartnic-survey-1}. In contrast, SCENIC provides comparable features and performance but as an open-source platform that combines offloaded network stacks, programmable congestion control, and GPU/SSD-Direct coupled with native Linux integration — a combination that no existing open research platform achieves — while remaining vendor-agnostic and fully accessible to the research community.

\subsection{Special-purpose research platforms}
Besides feature-complete NICs, both industry and academia have explored network
platforms that target specific aspects of the network rather than
providing a general-purpose host interface. AccelTCP~\cite{accelTCP} accelerates TCP management by offloading
connection states to ARM cores on a SmartNIC, bypassing the host kernel on the fast
path. iPipe~\cite{iPipe} further extends this approach and relocates distributed application
logic directly onto the NIC using an actor-based programming model. 
On FPGAs, a number of projects explore offloaded networking stacks. FlexTOE~\cite{flexTOE}, Limago~\cite{limago} and EasyNet~\cite{easyNet} propose full implementations of TCP/IP in FPGA fabric, while BALBOA~\cite{roce_balboa} proposes an RDMA stack with a focus
on datacenter compatibility. StRoM~\cite{stRoM} extends an
FPGA-based RoCEv2 stack with new opcodes for improved remote memory access in disaggregated memory systems.
ClickNP~\cite{clickNP} takes a more customizable approach, providing a modular
programming framework for composing high-throughput packet processing pipelines, while FlowBlaze~\cite{flow_blaze} focuses on hardware abstractions for stateful processing pipelines. On the transport level, ACCL+~\cite{accl+} implements collective communications on FPGAs, achieving performance comparable to software-based MPI, while FpgaNIC~\cite{fpgaNIC} proposes a networked FPGA platform with direct FPGA-to-GPU DMA. SuperNIC~\cite{superNIC} addresses multi-tenancy on FPGA-based SmartNICs by introducing dynamically scheduled network task chains that share FPGA fabric resources across tenants. A challenge in the design of SCENIC was to combine many of these ideas and incorporate them into a single, efficient design.

\begin{figure*}[t]
    \centering
    \includegraphics[width=0.85\linewidth]{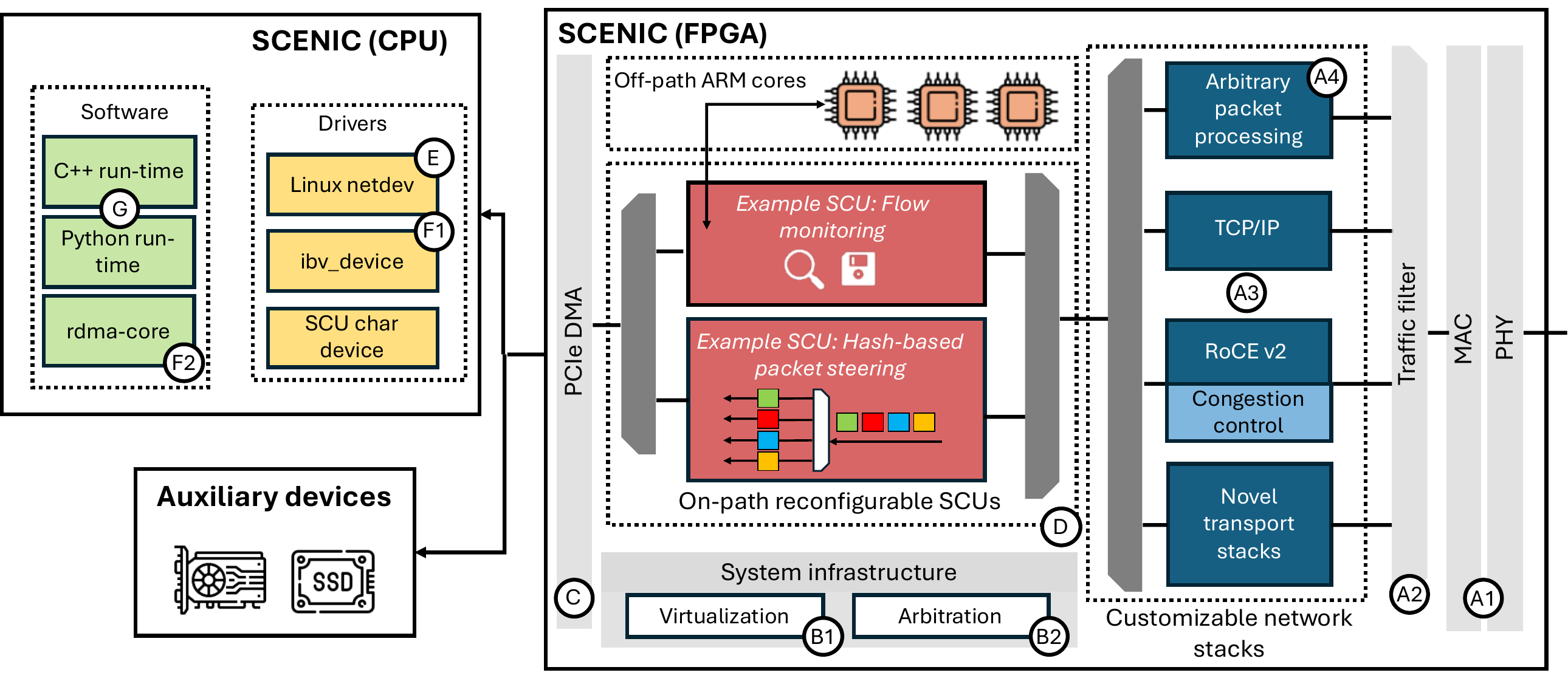}
    \caption{Overview of SCENIC with two example offloads: hash-based network-to-GPU data partitioning (Section~\ref{sec:gpu-hashing}) and hybrid flow monitoring (Section ~\ref{sec:arm-cores}).}
    \Description{Large overview diagram of SCENIC hardware, drivers and software.}
    \label{fig:scenic-overiew}
\end{figure*}

\section{Motivation \& Design Requirements}
Drawing on prior work and the characteristics of modern datacenters, we identify five requirements for our design:

\textbf{R1 – Performance:} Networking in the cloud and datacenters is rapidly shifting from 100G to 200G/400G~\cite{rdma_at_hyperscale, training_at_meta_scale}. Research SmartNICs, based on FPGAs, fail to achieve these bandwidths due to design complexity and clock frequency limitations. On the other hand, commercial SmartNICs achieve high bandwidth but rely on off-path ARM/RISC-V cores, which are unsuitable for latency-sensitive, high-throughput offloads~\cite{off_path_compute_1, off_path_compute_2}. SCENIC achieves the best of both worlds, achieving 200G with customizable, on-path streaming, and off-path ARM core offloads.

\textbf{R2 – Datacenter integration and compatibility:} A SmartNIC requires well-defined software and driver interfaces and support for standard transport protocols. Beyond that, modern workloads require direct interaction with heterogeneous GPUs~\cite{heterogeneous_gpu_datacenter} and storage~\cite{azure_rdma} at line rate. The challenge is doing so while also supporting in-network compute. SCENIC demonstrates both: compliance with existing software (exposed as Linux \texttt{netdev} and \texttt{ibv\_device}) and direct interoperability with GPUs and SSDs.

\textbf{R3 – Customization and programmability:} To utilize the available bandwidth, NIC customizability is key. As an example, DCQCN~\cite{dcqcn}, the congestion control algorithm hardwired into today's commodity RDMA NICs, is demonstrably suboptimal for modern traffic patterns~\cite{hpcc, revisting_cc_ethernet}. Yet, closed-source firmware and hardened hardware make it difficult to replace it or modify the transport layer to support novel semantics. SCENIC exposes the entire transport pipeline — from congestion control to the RDMA stack — as open-source FPGA IPs, enabling full customization and extensibility.
    
\textbf{R4 -- Support for multiple flows and isolation:} A production server hosts multiple tenants simultaneously, making isolation and fairness key requirements for a SmartNIC. However, many available SmartNICs often fall short of providing such guarantees due to the lack of virtualization and access control rules~\cite{smartnic-survey-1}. SCENIC addresses these through memory virtualization and flow arbitration, allowing the deployment of multiple, parallel and independent Stream Computation Units (SCUs). 

\textbf{R5 – Programming model:} Utilizing SmartNIC compute capabilities requires a user-exposed programming interface. In commercial SmartNICs, this typically mandates vendor lock-in via proprietary frameworks~\cite{smartnic-survey-1} such as DOCA~\cite{nvidia_doca_2024} or Pensando SSDK~\cite{amd_pensando_ssdk_2024}. Instead, SCENIC offers an open programming standard for offloads, supporting hardware description languages (HDLs), high-level synthesis (HLS), and network-specific languages such as P4.

\section{System overview}

SCENIC (Figure~\ref{fig:scenic-overiew}) consists of a modular hardware design, kernel-space drivers, and a user-space software API. In hardware, SCENIC implements a network datapath with IP cores for the MAC and PHY layers~\RobustCircled[0.5pt]{\scriptsize A1}, a traffic filter~\RobustCircled[0.5pt]{\scriptsize A2}, offloaded TCP/IP and RDMA stacks~\RobustCircled[0.5pt]{\scriptsize A3}, and a slow path for processing generic traffic~\RobustCircled[0.5pt]{\scriptsize A4}. Additional hardware components handle I/O virtualization~\RobustCircled[0.5pt]{\scriptsize B1} and fair resource sharing in multi-tenant systems~\RobustCircled[0.5pt]{\scriptsize B2}. Finally, the DMA engine~\RobustCircled[0.5pt]{\scriptsize C} is used for host-to-SCENIC data movement and interrupts. The offloaded applications, termed Stream Computation Units (SCUs)~\RobustCircled[0.5pt]{\scriptsize D}, can be used to implement custom functionality with access to all incoming and outgoing network traffic, as well as CPU/GPU memory and SSDs.

On the driver-side, SCENIC exposes a full Linux \texttt{netdev} \RobustCircled[0.5pt]{\scriptsize E} and implements a \texttt{rdma-core} verbs provider through a kernel-space driver~\RobustCircled[0.5pt]{\scriptsize F1} and standard user-space libraries~\RobustCircled[0.5pt]{\scriptsize F2}, exposing SCENIC as an RDMA-capable SmartNIC with support for standard IB Verbs. SCENIC further extends Coyote's driver stack with a host-side runtime~\RobustCircled[0.5pt]{\scriptsize G} for offload configuration and control, including a Python runtime compatible with standard libraries such as NumPy and Pandas.

SCENIC is designed with modularity in mind so that specific parts of the system can easily be enabled or disabled to produce application-specific designs. For example, workloads that do not require TCP/IP can disable it, freeing resources for SCUs. The only component always present is the slow path, ensuring Linux \texttt{netdev} compatibility. SCENIC's design targets datacenter workloads and includes the slow path, an offloaded RDMA stack, memory virtualization for CPU and GPU access, and one SCU. TCP/IP, multiple SCUs\footnote{SCENIC can be configured to include up to 16 independent SCUs.}, and access to SSD or FPGA memory (HBM/DDR) can be enabled via compile-time flags.

We prototype SCENIC on a range of widely available FPGAs: 100G UltraScale+ AMD Alveo devices (U55C, U280, U250) and the most recent Alveo V80 (Versal architecture). The V80 provides substantially higher bandwidth (200G+), integrates Arm cores, and features a Network-on-Chip (NoC), making it a significantly more powerful platform. However, unlike UltraScale+ devices, the V80 has no complete shell, requiring all low-level hardware blocks (e.g., PCIe DMA, memory controllers, and reconfiguration logic) to be implemented from scratch. 
SCENIC bases its design on Coyote v2~\cite{coyotev2}, an open-source shell designed for 100G UltraScale+ platforms, and open-source TCP/IP~\cite{fpga-network-stack-github} and RDMA~\cite{roce_balboa} stacks.
We redesigned many aspects of these projects to support higher throughput, direct GPU/SSD communication, and transparent software integration via \texttt{netdev} and \texttt{ibv\_device}. Additionally, we configured the 200G AMD DCMAC IP~\cite{dcmac-ip} to enable networking on the V80, for which no fully functional reference design existed. 
\section{Hardware architecture}

\subsection{Network datapath}
To sustain line-rate performance (\textbf{R1}), SCENIC implements the physical layer on top of the Xilinx CMAC (UltraScale+ platforms, 100G) or DCMAC (Versal platforms, 200G) IP cores. These manage PAM2/PAM4 signal modulation~\cite{PAM2, PAM4}, forward error correction (FEC), and priority flow control (PFC) for lossless Ethernet~\cite{pfc}. Because these MAC cores stream data in a bandwidth-dependent clock domain and do not support backpressure, we safely cross clock domains into the application logic using a buffered, deeply pipelined datapath in both the RX and TX directions. Beyond the MAC, the hardware mirrors the OSI model. A networking prefilter acts as a triage layer, separating the fast path from the slow path: TCP and RoCEv2 packets are routed to our hardware-offloaded networking stacks (if enabled), while all unhandled traffic is sent to the host via a dedicated DMA engine for Linux \texttt{netdev} processing (Section \ref{sec:linux_netdev}).

For the offloaded fast path, an ARP resolver and a frame header insertion module handle the data link layer, exposing L2-stripped, network-layer packets on an AXI Stream bus. A subsequent L3/L4 filter then dispatches packets to the appropriate transport stack. SCENIC relies on the same open-source network stack IPs as Coyote~\cite{roce_balboa, fpga-network-stack-github}. Crucially, these transport stacks decouple the control and data planes on the host-facing side. Extracted payloads are steered to specific Stream Computation Units (SCUs) based on control plane tags, such as the RoCE Queue Pair Number (QPN) or TCP Flow ID. This explicit, user-defined in-NIC routing enforces tenant and flow isolation (\textbf{R4}) and supports a fine-grained, per-packet programming model (\textbf{R5}).

\subsection{Congestion control and network extensions}
Large-scale ML workloads expose the limitations of fixed congestion control algorithms~\cite{hpcc, training_at_meta_scale}. Programmable congestion control (PCC), with scenario-adaptive algorithm selection, addresses this challenge but requires direct modification of the transport logic, which is restricted on commercial NICs. SCENIC’s open architecture removes this constraint, enabling a broad design space for customization (\textbf{R3}).

At the same time, any congestion control mechanism must satisfy the strict per-packet processing budget imposed by high link rates. At 200\,G with MTU-sized RoCE packets, this budget is approximately:
\[
  \frac{4178 \times 8\,\text{bit}}{200 \times 10^{9}\,\text{bit/s}} \approx 167\,\text{ns}
\]
This constraint effectively rules out embedded CPU-based approaches. For example, NVIDIA BlueField-3 DOCA PCC cannot operate on a per-packet basis at line rate, and achieving such performance requires fundamental architectural changes demonstrated only by recent research~\cite{swcc}.

\begin{figure}
    \centering
    \includegraphics[width=\linewidth]{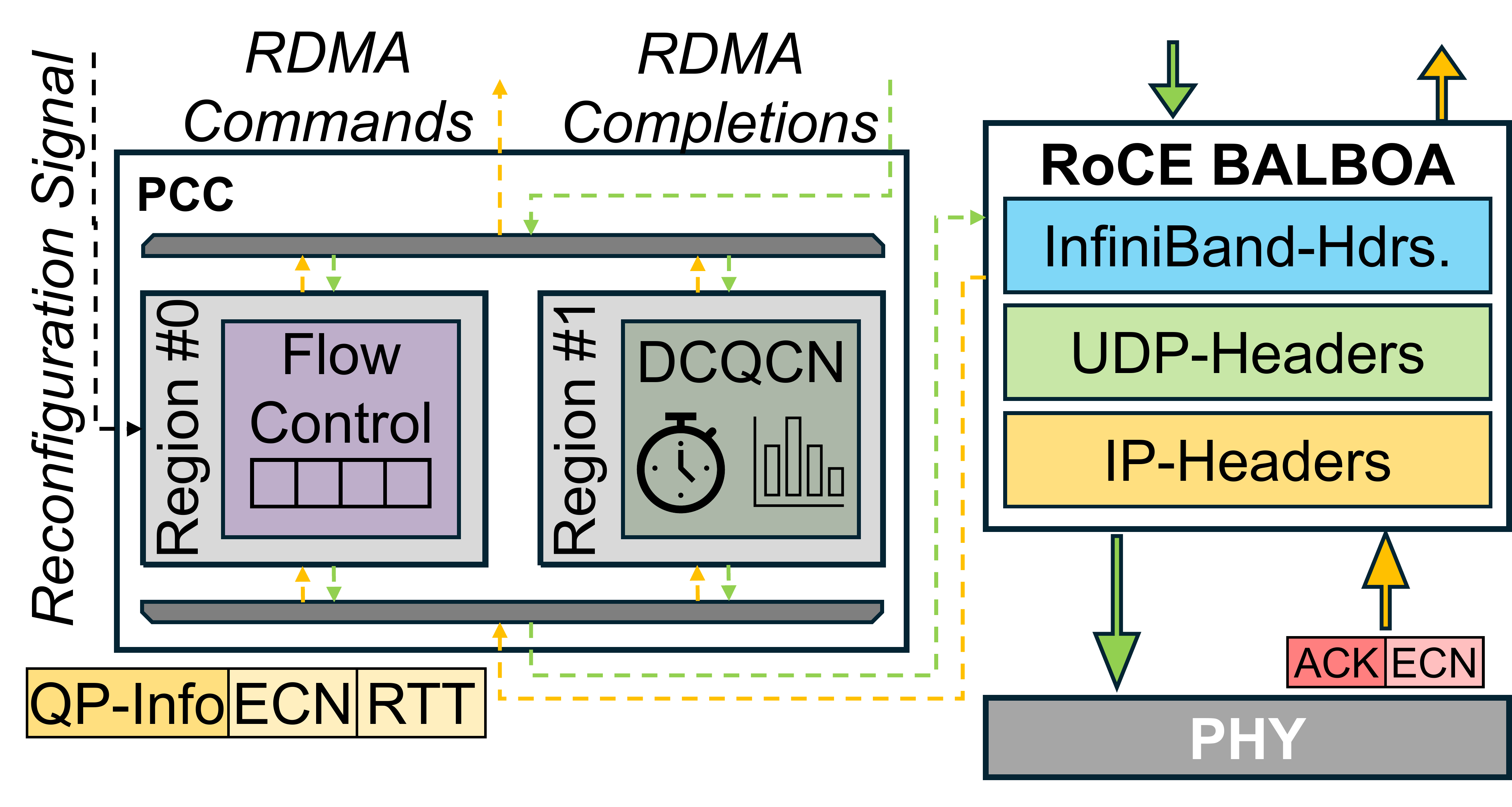}
    \caption{Programmable congestion control in SCENIC.}
    \label{fig:pcc_hardware}
\end{figure}

In contrast, FPGA-based pipelines provide deterministic processing, handling each packet in a fixed number of clock cycles. At 391\,MHz, 167\,ns corresponds to roughly 65 cycles, which is sufficient even for algorithms that process multiple telemetry signals per packet, such as SMaRTT~\cite{fastflow}. SCENIC leverages this property by implementing each congestion control algorithm as a dedicated hardware module, which can be swapped at runtime via dynamic partial reconfiguration of the FPGA fabric (\textbf{R3}). As reference implementations, SCENIC provides a simple ACK-clocked window-based flow controller and a full DCQCN implementation. In our experiments, we measure an average reconfiguration time of 8ms. This latency can be completely hidden through a dual-CC implementation as showcased in \autoref{fig:pcc_hardware}. While one CC-implementation is actively steering the command flow, a second preloaded algorithm is already receiving congestion signals. When reconfiguration is triggered, this second module immediately takes over congestion control. 

\subsection{I/O virtualization and isolation}
SCENIC extends Coyote's virtual memory model, which enables the FPGA to access host CPU memory and, more recently, AMD GPU memory. However, for SmartNIC use cases, these are not sufficient. We therefore further extend the model to support NVIDIA GPUs as well as NVMe devices (\textbf{R3}). Access to NVIDIA GPU memory follows the same approach as for AMD GPUs, using the Linux \texttt{dma-buf} mechanism. This mechanism allows PCIe devices (e.g., GPUs) to export memory regions that can be imported by other PCIe devices (e.g., FPGAs), and has recently become the standard implementation for GPUDirect RDMA~\cite{gpu-rdma}.

Beyond GPU memory, direct access to storage devices is equally critical for a SmartNIC that aims to minimize host CPU involvement on the I/O path. SCENIC therefore integrates a dedicated NVMe host controller within the FPGA fabric to handle both submission and completion logic. To bypass host CPU intervention, all NVMe control structures, including submission/completion queues and PRP lists, as well as data buffers are hosted in FPGA HBM/DDR and are mapped to a PCIe BAR. By maintaining direct access to NVMe doorbell registers and processing completion entries in hardware, SCENIC enables a low-latency data path between storage and network stacks.

Coyote's virtual memory model also ensures strict isolation (\textbf{R4}) between offloaded applications~\cite{coyotev2}. Resource fairness (\textbf{R4}) is achieved through system-wide arbiters, which ensure that all offloaded application equally share the available bandwidth through packet-based, round-robin arbitration. This enables SCENIC to natively support multiple RDMA QPs or TCP/IP sessions, each steered to a specific offload, with isolation and fairness guarantees.

\subsection{Host DMA engine}
SCENIC implements a low-latency, high-throughput DMA engine for NIC-to-host transfers and interrupts (\textbf{R1}). On 100G UltraScale+ platforms, DMA is realized through the XDMA IP~\cite{xdma-ip} supporting PCIe Gen3x16. On 200G Versal platforms, we use the hardened QDMA IP~\cite{qdma-ip}, which can be configured in PCIe Gen4x16 or Gen5x8 mode. Notably, the QDMA IP is designed for high-performance networking, with up to 2048 DMA queues. To maximize performance and reduce backpressure we configure the number of queues to match the number of outstanding packets. In both cases, like on commercial, ASIC-based NICs, we explicitly enable relaxed ordering for reads, reducing latency and preventing head-of-line transfer stalls. The same DMA engines are used to implement MSI-X interrupts between the SmartNIC and the host CPU, as required for \texttt{netdev} integration and other sources of interrupts. However, when processing interrupts, care has to be taken with parallelism: by default, multiple interrupt workers can be launched in parallel, which, while maximizing performance, can also lead to race conditions (e.g., the same interrupt would be launched for two packets arriving in a short amount of time). To prevent such cases, the SCENIC driver (Section~\ref{sec:linux_netdev}) implements various locks and mutexes for shared resources, while also limiting the number of interrupt threads where applicable.

\section{Application offloads}

\subsection{Streaming on-datapath offloads}
SCENIC's SCUs implement custom user logic and can access host CPU, GPU, and SSD memory, as well as all incoming and outgoing network traffic. This enables streaming, dataflow-style designs commonly used in network processing tasks such as compression, encryption, and hashing. What sets the SCUs apart from offloads on conventional SmartNICs is the combination of compute flexibility and memory bandwidth: FPGAs provide fine-grained control over parallelism, hardware primitives, and accesses patterns. While most SmartNICs rely on DRAM with tens of GB/s throughput, SCENIC supports HBM on platforms such as Alveo V80, U55C, and U280 with hundreds of GB/s of memory bandwidth. On other platforms (e.g., Alveo U250), it leverages on-board DRAM, while all supported FPGAs include tens of megabytes of on-chip Block- and Ultra-RAM (BRAM/URAM) with single-cycle access latency. This heterogeneous memory hierarchy, combined with fine-grained hardware control, enables high-throughput, latency-sensitive data processing (\textbf{R1}).

Equally important is the SCU programming model (\textbf{R5}). As a network peripheral in the datacenter, SCENIC must allow SCUs to be developed quickly without requiring extensive hardware design knowledge. Thus, SCUs in SCENIC can be implemented in one of the following ways:
\begin{itemize}
    \item \textbf{Register-transfer level (RTL),} through languages such as SystemVerilog and VHDL. These languages give full control over the underlying hardware and best performance, but require considerable hardware expertise and long development cycles.
    \item \textbf{High Level Synthesis (HLS)}, a programming language based on C++ with pragmas to guide the hardware behavior. Developing with HLS is far more accessible, leading to up to 75\% reduced development times, but often at the expense of resource consumption~\cite{hls-survey}.
    \item \textbf{SpinalHDL}, a Scala-based hardware construction language that generates synthesizable RTL~\cite{spinalhdl-doc}. It provides higher abstraction than traditional RTL through parameterization, object-oriented, and functional programming, improving code reuse and productivity while preserving full hardware control.
    \item \textbf{P4}, a language for packet processing pipelines, enabling rapid development of parsing, classification, and header manipulation. It is limited to match-action style processing and less suited for complex computations. Integration is done via AMD Vitis P4 IP~\cite{p4-ip}, which synthesizes P4 code into FPGA RTL.
\end{itemize}

P4 SCUs are best suited for the slow path, which processes full headers, while HDL- and HLS-based SCUs are better suited for complex computational tasks where the header is stripped by the network stack (e.g., RDMA). In all cases, SCENIC's build flow automatically compiles and synthesizes SCU code and links it with the rest of the platform. SCENIC also natively integrates with an open-source library of FPGA components\footnote{https://github.com/fpgasystems/libstf}, providing commonly used building blocks such as stream normalizers, hashing modules, crossbars, and data width converters. For AI workloads, SCENIC additionally integrates with ACCL+~\cite{accl+}, a collective offload engine for FPGAs with performance comparable to MPI. Finally, SCENIC incorporates an extensive simulation environment, enabling SCUs to be verified before deploying on hardware.

\subsection{ARM-based off-datapath offloads}
\label{sec:arm-cores}

While existing literature points to the limitations of off-path Arm cores as primary means for data processing on SmartNICs~\cite{off_path_compute_1, off_path_compute_2}, SCENIC demonstrates how such cores can effectively complement the on-path SCUs for control plane operations. While dynamic SCU reconfiguration enables flexible data path adaptation, it also requires taking a flow offline during reconfiguration. Control plane updates, such as security rules or telemetry tasks, can instead be handled on an off-path core without interrupting the flow, satisfying the customizability requirement (\textbf{R3}). Additionally, the ability to run standard software on these cores provides a lightweight and accessible programming model (\textbf{R5}).

On the V80 platform, SCENIC utilizes the embedded processing subsystem, consisting of a dual-core Arm® Cortex®-A72 application processor and a dual-core Arm® Cortex®-R5F real-time processor. Two types of interfaces enable tight coupling between the SCUs and the Arm cores: up to 16 IRQ connections can be dynamically assigned to individual SCUs providing low-latency event signaling from the data path to the Arm cores. Data transfer between the SCUs and the Arm cores is handled via a memory-mapped AXI bus, providing access to either control registers for lightweight command exchange or a dedicated memory unit for higher-capacity data buffering. The measured latency of this interface for a single access is about 0.3 $\mu$s, and the single-trip time of an SCU interrupt reaching the handler in the Arm core is about 0.2 $\mu s$ — sufficient for transferring aggregated statistics, flow state snapshots, or control instructions to the processor cores without becoming a bottleneck for control plane operations. 

We demonstrate this SCU–CPU co-design with an incast traffic firewall. Stateful firewalls at 100G+ line rate exceed CPU capabilities, while FPGA-only solutions lack policy flexibility, requiring convoluted state machines~\cite{flow_blaze}. SCENIC resolves this by offloading line-rate flow tracking to an SCU pipeline, classifying flows by source subnet to reflect pod-level positions in Fat Tree topologies~\cite{fat-tree}, while implementation dynamic policy decision making on the Arm cores. A hardware timer periodically interrupts the CPU, which reads traffic statistics via the AXI bus. A dynamically configurable SCU rate limiter then enforces the resulting policies. 

\section{Driver and software integration}
In the following, we present SCENIC's software API, as well as its native integration with Linux \texttt{netdev} and \texttt{rdma-core} to ensure compatibility with existing applications (\textbf{R2}).

\subsection{Linux \texttt{netdev} integration}
\label{sec:linux_netdev}
\begin{figure}
    \centering
    \includegraphics[width=\linewidth]{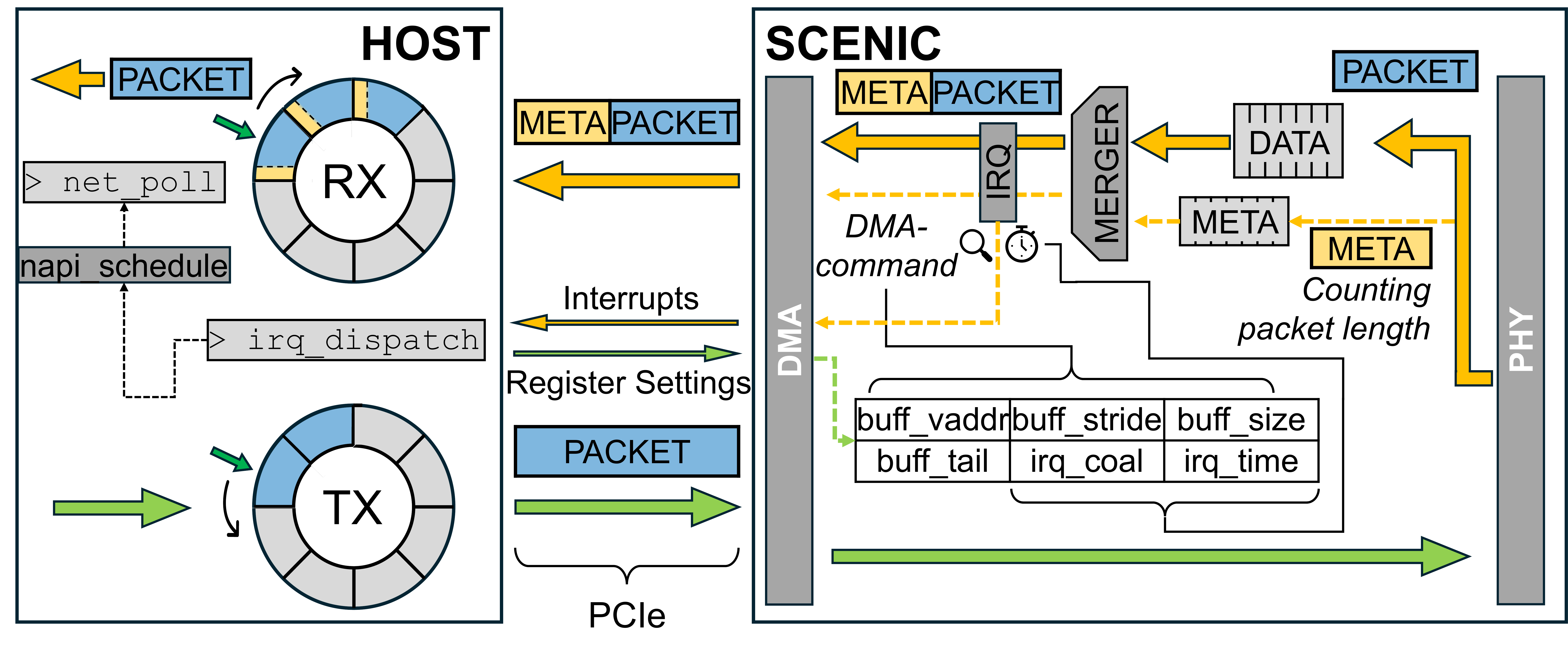}
    \caption{DMA packet forwarding to the network driver.}
    \label{fig:APP_DMA_logic}
\end{figure}

\autoref{fig:APP_DMA_logic} visualizes SCENIC's arbitrary packet processing path which uses a hardware-software co-design to minimize DMA overhead and interrupt cost. On the RX path, a hardware state machine prepends a compact metadata tag, containing the packet length and a valid flag, directly before each packet payload. This allows the tag and the payload to share a single DMA transaction into a unified ring buffer, halving the number of required DMA operations compared to a naive two-transfer design and resulting in improved performance (\textbf{R1}). The driver's \texttt{net\_poll} handler then reads each metadata tag to determine packet validity and length before copying the payload starting right after the tag to the host networking stack.
Interrupt delivery is managed by an MSI-X controller with two tunable behaviors: a timeout threshold guarantees bounded latency under sparse traffic, while interrupt coalescing amortizes interrupt overhead under high load, improving throughput. Both parameters are configurable by the driver, allowing the tradeoff to be tuned per workload. 
The TX path mirrors this structure of the RX path, using a ring buffer to enqueue outgoing commands.

In terms of offloaded network acceleration, the blocks used for the physical network layer (CMAC and DCMAC) implement the Ethernet checksum calculation. Further features such as TSO or GRO are not implemented in hardware to keep the design small and resource-efficient, but can be added in future work.

\subsection{\texttt{ibv\_device} integration}
For data-intensive workloads, SCENIC includes a fully offloaded RoCEv2 stack and exposes itself as an \texttt{ibv\_device}, ensuring compatibility with existing IB Verb applications (\textbf{R2}). This is realized through a two-component stack: (1) a kernel driver manages low-level hardware setup, including mapping of control registers and communication of link properties to the operating system, and (2) an ABI-defined interface connecting the driver to the \texttt{scenic\_ib} provider implementation in the \texttt{rdma-core} userspace library. Accessing the mapped writeback and control registers allows the provider to interact directly with SCENIC's DMA engine for outgoing \texttt{RDMA WRITEs} and \texttt{RDMA READ REQUESTs} and completion checking.

Three design points illustrate the hardware-software co-design principle concretely. When calling \texttt{ibv\_reg\_mr}, the provider stores memory translations for the allocated user-space buffer directly in SCENIC's on-device Translation Lookaside Buffers (TLBs). SCENIC's driver ensures scalability across hundreds of QPs through automated translation lookups on TLB misses, while an LRU replacement strategy implemented in hardware preserves low latency for frequently active QPs (\textbf{R1}). Similarly, SCENIC's completion mechanism avoids interrupt overhead entirely: the hardware performs atomic increments of per-QP completion counters, which the \texttt{rdma-core} provider polls directly, yielding high throughput without interrupt processing overhead. Finally, we implement \texttt{ibv\_create\_qp\_ex} with an SCU index as an
additional parameter, allowing applications to explicitly map RDMA flows to
specific SCU processing pipelines~(\textbf{R3}). Since each SCU maintains
dedicated, non-shared datapath resources, this mapping provides hardware-level
isolation between Queue Pairs assigned to different SCUs~(\textbf{R4}).

\begin{figure}[t]
    \centering
    \includegraphics[width=\linewidth]{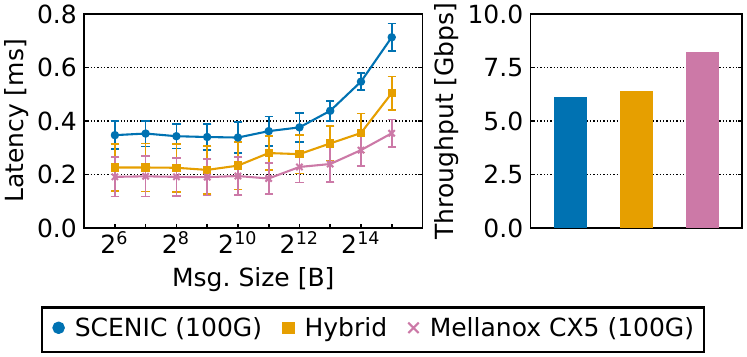}
    \caption{Performance evaluation of the fallback path. Left: \texttt{ping} latency. Right: \texttt{iperf3} throughput. Hybrid refers to Mellanox-to-SCENIC communication. }
    \label{fig:host_networking_performance}
\end{figure}

\subsection{Software API}
To control and configure the SCUs, SCENIC implements a high-level (\textbf{R5}) software run-time in both Python and C++, based on Coyote's software API. The software run-time allows various SCU-related control tasks, including setting and reading control registers, polling for interrupts, data movement and dynamic reconfiguration. Of particular interest is SCENIC's Python run-time, which enables high-level control compatible with modern data processing libraries (e.g., PyTorch, Pandas). To ensure high performance, the Python run-time is implemented as a thin wrapper around the C++ run-time, thus minimizing overhead. An example of dynamically loading the application, setting control registers and checking completions is shown below.

\begin{code}
\captionof{listing}{Example Python code for SCU control}
\begin{minted}{python} 
# Load target application to SCU 2
ScenicReconfig().reconfigure_app(
    2, "/path/to/scu/bitstream"
)
 
# Create a thread and assign it to SCU 2
scenic_thread = ScenicThread(2)

# Set control register, e.g., encryption key
scenic_thread.set_csr(0x9f3c7a2b6e41d8c5, 0);

# Check completions, for e.g., RDMA WRITE
scenic_thread.get_completed("remote_write")
\end{minted}
\end{code}

\section{Evaluation}
To evaluate SCENIC's performance, we conduct microbenchmarks of the various components: slow-path packet processing, RDMA offload engine, GPU and SSD integration, and flow isolation. For a better understanding of the measurements, we provide comparisons with standard datacenter NICs, 100G Mellanox ConnectX-5 and 200G Broadcom. We evaluate SCENIC on the AMD Alveo U55C for 100G designs and on the AMD Alveo V80 for 200G designs. The measurements are conducted in a public research cluster, the AMD-ETH Heterogeneous Compute Cluster~\cite{hacc}, which uses fully switched 100G and 200G networks.

\begin{figure*}[t]
    \centering
    \includegraphics[width=\linewidth]{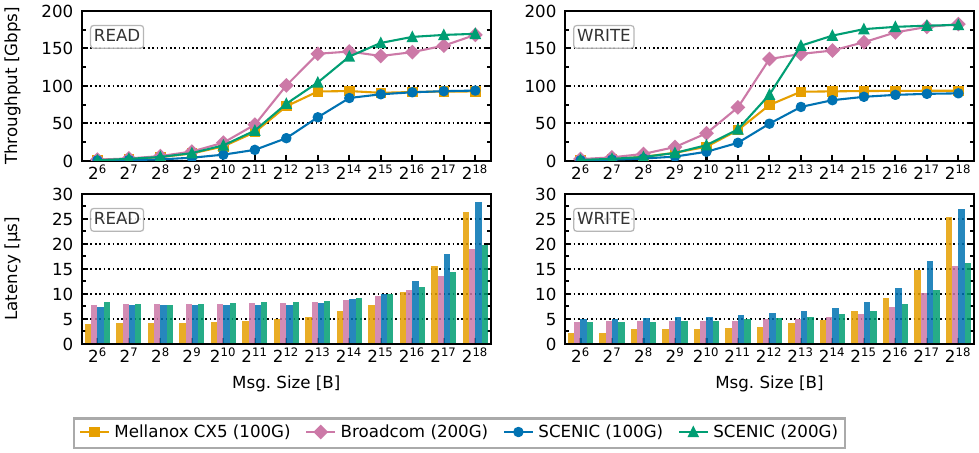}
    \caption{RDMA performance benchmark in a fully switched datacenter network.}
    \label{fig:rdma_microbenchmark}
\end{figure*}

\subsection{Host networking performance}
Due to the integration with \texttt{netdev}, we can use standard Linux tools to evaluate the performance of the slow path and the network driver for packets that cannot be processed by any of the offloaded networking stacks. To match the network offload found in SCENIC's fallback path and measure the general datapath and driver performance, on-NIC segmentation offloads (TSO and GRO) are turned off on the commercial NICs.
The average latency and jitter between SCENIC and commercial NICs through the \texttt{ping} utility (ICMP request and response) over a 100G link is shown in \autoref{fig:host_networking_performance}. Commercial, ASIC-based NICs likely benefit from generally higher clock speeds and further optimized DMA engines, resulting in slightly lower latencies. Despite this gap, SCENIC's slow-path latency is well within the range required for management traffic: SSH sessions, monitoring, and control-plane communication remain fully responsive under all tested conditions.
A similar conclusion can be drawn for achievable throughput with \texttt{iperf3} at a typical MTU of 1500B and no further optimizations. While falling slightly behind commercial NICs, the performance is sufficient for practical tasks like video streaming or scp-based data transfer. This is expected and acceptable by design since bulk data transfers are handled at line rate by the networking stacks.

\subsection{RDMA performance}
SCENIC's full exposure as \texttt{ibv\_device} enables the standard \texttt{perftest} benchmark~\cite{perftest} to be run without modifications, providing a direct comparison to commercial NICs (\autoref{fig:rdma_microbenchmark}). 
Latency measurements via \texttt{ib\_write\_lat} and \texttt{ib\_read\_lat} show a slight advantage for the Mellanox, consistent with a more mature PCIe DMA engine and the higher clock speed of the ASIC NIC. However, this gap is modest and does not impact SCENIC’s target use cases, for which throughput and programmability represent the main objectives. For throughput, SCENIC saturates the available network bandwidth for both \texttt{ib\_write\_bw} and \texttt{ib\_read\_bw} and achieves performance comparable to commercial NICs. 

\subsection{GPU integration}
We measure end-to-end throughput between SCENIC and GPUs using RDMA READs and WRITEs. In this configuration, the GPU node acts as the responder (server). For READ benchmarks, the client reads data from GPU memory and polls for local delivery; for WRITE benchmarks, the requester writes data to the server's GPU memory and polls on acknowledgments generated by the server-side NIC. Important to note, NIC-side acknowledgments are generated independently from the GPU memory controller; however, we confirm similar throughput by also polling on memory contents of the server-side GPU.

\begin{figure}[b]
    \centering
    \includegraphics[width=0.9\columnwidth]{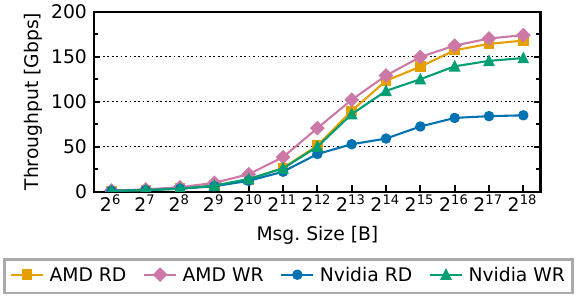}
    \caption{Throughput of SCENIC to GPU with RDMA READs and RDMA WRITEs.}
    \label{fig:gpu_results}
\end{figure}

As shown in Figure~\ref{fig:gpu_results}, performance on AMD GPUs saturates the link and is comparable to our CPU baseline (Figure~\ref{fig:rdma_microbenchmark}). In both cases, we observe lower performance on NVIDIA GPUs. Part of this can be attributed to a sub-optimal PCIe configuration of the NVIDIA node, over which we have limited control. We confirm the sub-optimal performance by measuring local (CPU-to-SCENIC) transfers and observe saturation at 20 GBps, rather than the expected 23 - 25 GBps for a PCIe Gen4x16 set-up. Additionally, READs exhibit higher performance degradation on NVIDIA GPUs and warrant further investigation. However, prior works, on both AMD~\cite{ropeerto} and NVIDIA~\cite{zero_nic} GPUs have noted lower performance for GPU-FPGA transfers, especially for READs. We plan on addressing these bottlenecks in future, but the presented results showcase SCENIC's capability for vendor-agnostic integration with GPUs.

\subsection{SSD integration over TCP/IP}
We evaluate SCENIC's TCP/IP offload engine on a network-to-storage path. On the U55C platform, we construct a pipeline in which incoming TCP/IP segments are reassembled in hardware and written directly into a local NVMe SSD, fully bypassing the host CPU. We compare against a host baseline using a 100G Mellanox ConnectX-5, where incoming data is received through the kernel TCP stack and written to the SSD via a single-threaded \texttt{io\_uring} event loop with \texttt{O\_DIRECT}. 

\begin{figure}[bt]
    \centering
    \includegraphics[width=\linewidth]{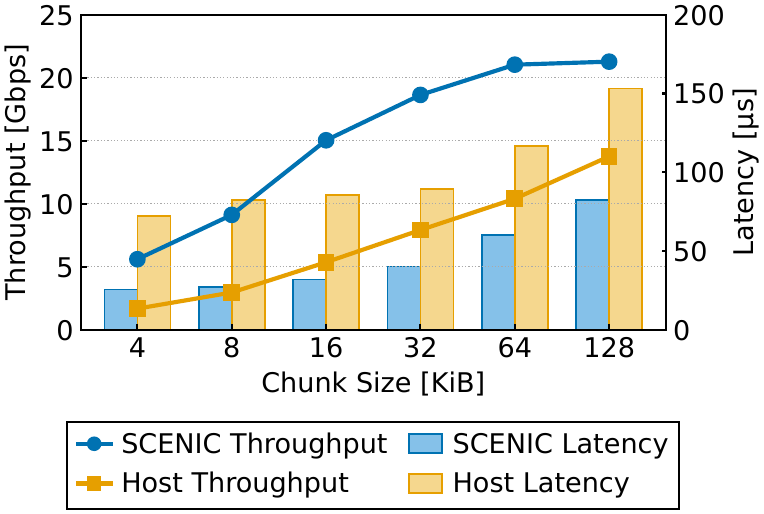}
    \caption{TCP to NVMe throughput and latency with SCENIC offload, compared to the host-side software baseline.}
    \label{fig:nvm_results}
\end{figure}

Figure~\ref{fig:nvm_results} compares the latency and throughput of both configurations. SCENIC achieves 2--3$\times$ lower per-completion latency across all chunk sizes (e.g., 25.6\,$\mu$s vs.\ 72.4\,$\mu$s at 4\,KB). A breakdown of the end-to-end latency reveals that software overheads, including kernel TCP/IP packet processing, system call transitions, and buffer management, dominate the host path (56--115\,$\mu$s), whereas SCENIC's hardware path only adds 14--27\,$\mu$s. 
In terms of throughput, even with 8 outstanding requests, SCENIC saturates the NVMe bandwidth and outperforms the host's best configuration with 64 outstanding requests by 1.27--1.47$\times$ across all chunk sizes. Moreover, since SCENIC entirely bypasses the host CPU, the architecture is expected to scale to multiple SSDs without the software overhead.

\subsection{Performance isolation through separate SCUs}
\begin{figure}[bt]
    \centering
    \includegraphics[width=\linewidth]{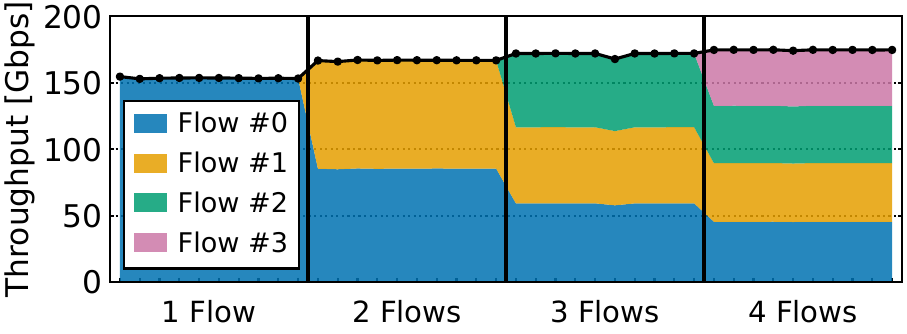}
    \caption{Time series of bandwidth sharing scaling up to four parallel flows performing 128 KiB RDMA READs through separated SCUs.}
    \label{fig:isolation_stacked_throughput}
\end{figure}

The strict, hardware-level separation of network flows through different SCUs does not only allow fine-grained, per-flow processing, but also guarantees fairness through arbitration (\textbf{R4}). To demonstrate this design aspect, we configure SCENIC with four SCUs and run a throughput benchmark with 128 KiB \texttt{RDMA READs}. We incrementally scale the workload from one to four parallel flows, mapping each distinct flow to its own SCU. \autoref{fig:isolation_stacked_throughput} shows a time series of the throughput as more flows are added to the system. As required, the aggregate throughput saturates the expected READ-bandwidth over a 200G link while being equally shared among active flows, even with new ones being added. This demonstrates that the SCU-based architecture prevents cross-flow interference, ensuring that independent network streams do not contend for bandwidth-constrained resources under full load.

\subsection{Resource consumption}
The resource consumption of the core SCENIC design, consisting of the slow path, an offloaded RDMA stack, the memory virtualization layer, one SCU and the host DMA engine is shown in Table~\ref{tab:resources}. SCENIC's resource consumption remains low: less than 30\% on the U55C/U250 and less than 15\% on the V80, leaving ample room for offloaded applications.

\begin{table}[b]
\caption{SCENIC resource consumption.}
\label{tab:resources}
\begin{tabular}{l|lll}
Platform & LUT {[}\%{]} & REG {[}\%{]} & BRAM {[}\%{]} \\ \hline
V80      & 11.5         & 12.5         & 17.1          \\
U55C     & 28.8         & 25.8         & 25.2          \\
U250     & 22.5         & 19.7         & 18.8              
\end{tabular}
\end{table}

\section{Use cases}

\subsection{ACCL}
\begin{figure}[t]
    \centering
    \includegraphics[width=\columnwidth]{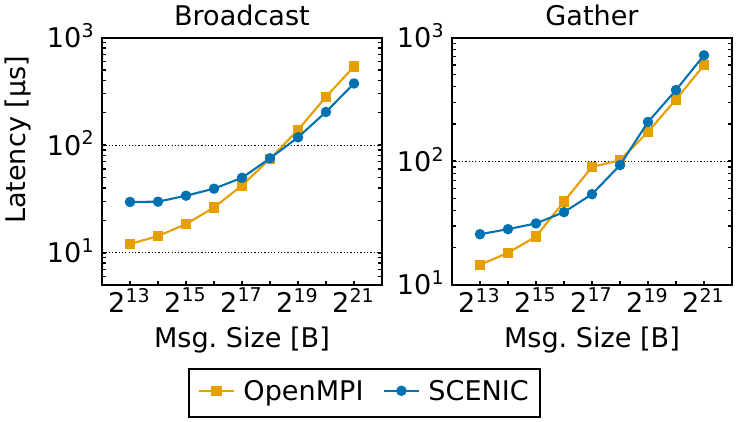}
    \caption{Comparison of \texttt{BROADCAST} and \texttt{GATHER} collectives on SCENIC with OpenMPI on a commercial NIC.}
    \label{fig:accl_perf}
\end{figure}

As a first use case, we deploy the open-source ACCL+~\cite{accl+, xilinx_accl_github} library on SCENIC, enabling offloaded collective communication. \autoref{fig:accl_perf} shows the results of a four-node benchmark comparing the \texttt{BROADCAST} and \texttt{GATHER} collectives on SCENIC with ACCL+ against the CPU baseline with RDMA OpenMPI and a Mellanox Connect-X5. SCENIC matches the performance of the commercial NIC with established libraries, while providing two clear advantages. First, collective communication represents a large part of the datacenter tax and the network bottleneck in large-scale AI training~\cite{megascale, FPGA_Smart_NIC_Scalable_AI}; offloading them to the network can free up CPU cycles or reduce the GPU utilization~\cite{nccl_sharp}. Second, offloaded collectives create the possibility of collocating
gradient compression as an in-network processing step to further overlap compute and communication~\cite{gradient_compression_smartnic}. For future work we plan to extend this example into an end-to-end pipeline for large-scale model training on GPUs with SmartNIC-offloaded collectives and compression.

\subsection{Hash-based data partitioning}
\label{sec:gpu-hashing}
As a second use case, we demonstrate SCENIC as an accelerator for cloud-native data processing systems. 
SmartNICs have emerged as a compelling platform for line-rate data decoding and parsing~\cite{conf/damon/DannW0FF22}, as well as for offloading data scanning and filtering~\cite{journals/corr/abs-2602-18775}---operations that impose significant runtime and memory overheads on host CPUs.
With this application, we target hash partitioning, a critical building block for multi-GPU query execution \cite{journals/pvldb/KabicWDA25} in modern data processing systems.
Hash partitioning is used to deterministically split up data into equal parts that can be processed independently  by, e.g., a join or aggregation operator.

We implement hash partitioning as a SCENIC SCU.
The SCU maintains an on-chip hash buffer ($16 \times 2^{16}$ hashes in our configuration) which supports hash folding over composite key columns.
The resulting hashes are then used to partition a set of data columns.
To support data sets that exceed the buffer capacity ($>2^{19}$ rows), we use batching.
For each GPU in the system, a dedicated pipeline selects the payload rows for each data columns and accumulates them in an output buffer, which is flushed in transfers of $64$kB (the smallest transfer size that saturates PCIe throughput).

\begin{figure}[t]
    \centering
    \includegraphics[width=\columnwidth]{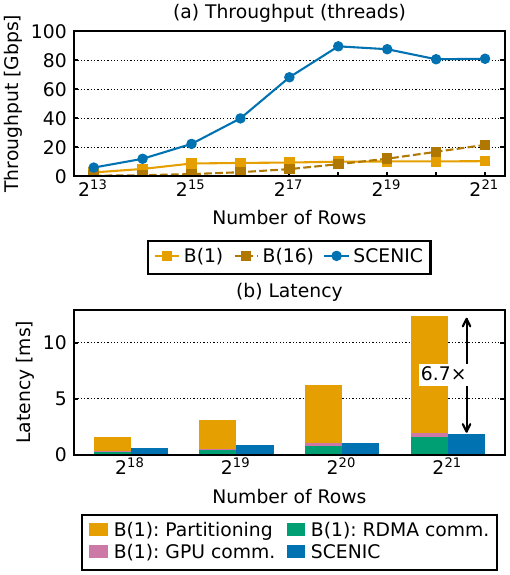}
    \caption{Performance of hash partitioning on the CPU (B: Baseline, 1 and 16 threads) and offloaded with SCENIC.}
    \label{fig:hash_part}
\end{figure}

We evaluate this use case on a two-node setup: a data processing node (AMD U55C and 4 AMD MI210 GPUs) and a remote memory node (AMD U55C). 
We generate a synthetic two-column table---one key column for hashing and one data column as the payload that is distributed across GPUs. 
Figure \ref{fig:hash_part} shows the throughput and the latency for the multi-threaded software baseline compared to SCENIC-offloaded hash partitioning. 
SCENIC offloading achieves latency that scales linearly with data set size and, for larger transfer sizes, approaches the lower bound of just the RDMA communication.
Throughput shows a fixed overhead at small data set sizes that is amortized for larger data sets.
Performance drops slightly beyond the on-chip hash buffer capacity, where batching is required.
The software baseline incurs substantially higher latency and reaches lower throughput, with thread management overhead limiting scalability at small transfer sizes. This use case demonstrates one of the many possibilities offered by SCENIC to accelerate applications and move data to where it can be best executed, thereby improving the utilization of expensive accelerators. 

\section{Conclusions and future work}
In this paper, we presented SCENIC, a fully open-source, 
200G, FPGA-based SmartNIC suitable for deployment in heterogeneous compute environments with CPUs, GPUs and SSDs. SCENIC demonstrates several innovative aspects in FPGA design for scalability, compatibility across architectures, and flexibility in network support for different protocols and features. It does this while being a full featured NIC rather than just focused on a single function, as most research prototypes do. Due to its native integration with Linux \texttt{netdev} and \texttt{rdma-core}, SCENIC works out of the box with existing applications, while providing easy-to-use Python and C++ run-times for offload configuration and control. Reconfigurable SCUs and off-datapath Arm cores enable low-latency, line-rate dataflow offloads, as shown by hash-based network-to-GPU data partitioning, achieving a $6.7\times$ latency improvement over the baseline.

In the future, we plan to further improve SCENIC in a variety of ways, with a primary emphasis on performance scaling to 2x200G with bifurcated PCIe Gen5x16, leading to up to 400G throughput. Additionally, we plan to explore hybrid application offloads, leveraging both the off-path ARM cores and the SCUs with suitable workload partitioning between the two. Finally, future work will involve new network protocols (e.g., UltraEthernet) and congestion control algorithms with SCENIC as a starting point.

\begin{acks}
We would like to thank AMD for the donation of the Heterogeneous Accelerated Compute Cluster (HACC) at ETHZ which was
used for the development of this project and Geert Roks for the support and help with the cluster set-up. This work was funded in part through an unrestricted grant from AMD. Additionally, we would like to thank contributors of the SLASH project from AMD Research Dublin, in particular Alexandru Ulmămei, Lucian Petrica and Mario Ruiz, for insightful guidance and help with the V80 FPGA. We have used Google Gemini and Anthropic Claude for code debugging, language checking, as well as for the homogenization of the plotting scripts used to generate the diagrams.
\end{acks}

\bibliographystyle{ACM-Reference-Format}
\bibliography{bibliography}

\end{document}